\documentclass[preprint,12pt,authoryear]{elsarticle}

\usepackage{amssymb}
\usepackage{wrapfig}
\setcitestyle{numbers,square}
\usepackage{float}
\usepackage{tikz}

\usepackage{graphicx}
\graphicspath{{.}{./Image/}}

\usepackage[authoryear]{natbib}
\setcitestyle{authoryear, open={[}, close={]}}
\usepackage{subcaption} % 导入子图包
\journal{Nuclear Physics B}

\begin{document}

\begin{frontmatter}



\title{\textbf{Patent Value Characterization — An Empirical Analysis of Elevator Industry Patents}}


\author{\textbf{Yuhang Guan, Runzheng Wang, Lei Fu, and Huane Zhang}}

% \address{School of Computer Science and Technology, Shandong University, Qingdao 266237, China}

% \affiliation{organization={School of Computer Science and Technology},
%             addressline={Shandong University}, 
%             city={Qingdao},
%             postcode={266237}, 
%             % state={},
%             country={China}}
\begin{abstract}

The global patent application count has steadily increased, achieving eight consecutive years of growth.The global patent industry has shown a general trend of expansion. This is attributed to the increasing innovation activities, particularly in the fields of technology, healthcare, and biotechnology. Some emerging market countries, such as China and India, have experienced significant growth in the patent domain, becoming important participants in global patent activities.

\end{abstract}


% \begin{highlights}
% \end{highlights}

% \begin{keyword}
% %% keywords here, in the form: keyword \sep keyword

% %% PACS codes here, in the form: \PACS code \sep code

% %% MSC codes here, in the form: \MSC code \sep code
% %% or \MSC[2008] code \sep code (2000 is the default)

% \end{keyword}

\end{frontmatter}


\section{Introduction}
\label{sect:Introduction}
\par
%由WIPO专利数据统计，从2009年开始，全球专利申请数量稳定上升，实现了连续8年增长。全球专利行业的规模普遍呈现增长趋势。这是由于创新活动的增加，尤其是在科技、医疗和生物技术领域。一些新兴市场国家如中国和印度在专利领域取得了显著的增长，成为全球专利活动的重要参与者。
%专利的落实需要进行融资，专利融资机构提供贷款时，首先需要考虑各种专利指标来估算其价值，然后根据专利的价值确定贷款金额。过去，专利价值的估算依赖于粗糙的数据统计和专利融资机构的主观判断。然而，在处理信息不足或包含过多抽象信息的专利时，这些方法变得不够有效。不合理的贷款可能导致机构财务问题，并对未来的专利贷款产生负面影响。这给专利融资带来了挑战。
%目前在对专利价值的研究中，对如何构建专利评估指标体系的研究最多，目前研究的重点是使用AI模型分析和传统数据分析的方式分析专利价值。本文以电梯领域为案例，对专利的各种定量和分类标签进行特征化和相关性分析，构建了一个专利评价指标体系。除此之外，我们使用决策树分类模型构建了一个专利价值分类预测模型，叫PVCP(patent value classification prediction) model，以预测专利的价值。这一预测将为专利融资机构提供指导。
%我们的主要贡献总结如下：
%1.我们构建了一个专利评价指标体系。
%2.我们为专利融资机构提供了科学可靠的专利估值参考。
\begin{wrapfigure}{r}{0.5\textwidth}
  \centering
  \includegraphics[width=\linewidth]{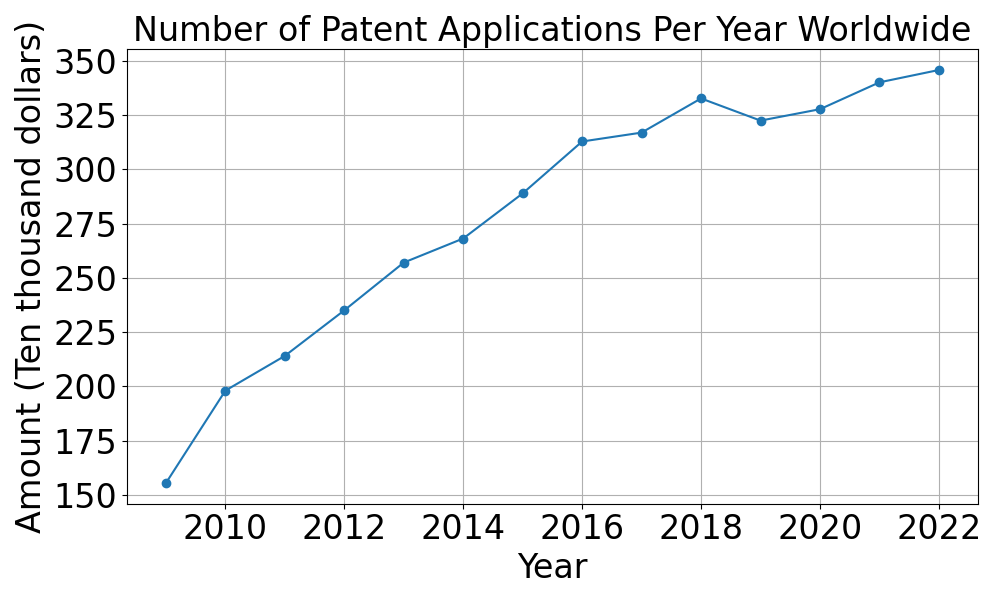}
  \caption{The Global Annual Number of Patent Applications}
  \label{fig:num_year_analysis}
\end{wrapfigure}
According to WIPO patent data statistics\footnote{Source: https://www.wipo.int/pressroom/en/}, starting from 2009, the global patent application count has steadily increased, achieving eight consecutive years of growth.The global patent industry has shown a general trend of expansion. This is attributed to the increasing innovation activities, particularly in the fields of technology, healthcare, and biotechnology. Some emerging market countries, such as China and India, have experienced significant growth in the patent domain, becoming important participants in global patent activities.\par
% With the development of science and technology, the number of patents in China has been steadily increasing year by year. According to data released by the Chinese government's official website, as of the end of 2022, the effective number of invention patents in China has reached 4.212 million. The realization of patents often requires substantial financial resources. With such a large number of patents, it presents both opportunities and challenges to patent financing institutions. For these reasons, the establishment of a scientific patent valuation criteria is of utmost importance.\par
The implementation of patents requires financing. When patent financing institutions provide loans, they first need to consider various patent indicators to estimate their value. Subsequently, the loan amount is determined based on the assessed value of the patents. In the past, the estimation of patent value relied on rough data statistics and subjective judgments of patent financing institutions. However, these methods become less effective when dealing with patents with insufficient information or containing excessive abstract information. Irrational loans may lead to financial issues for institutions and have a negative impact on future patent financing. This poses challenges to patent financing.\par
Currently, in the research on patent value, the focus is primarily on how to construct a patent evaluation indicator system. The current research emphasis is on analyzing patent value using AI models and traditional data analysis methods. This article takes the elevator industry as a case study, characterizes various quantitative and categorical labels of patents through feature extraction and correlation analysis, and constructs a patent evaluation indicator system. Additionally, we use a decision tree classification(DTC\citep{quinlan1986induction}) model to build a Patent Value Classification Prediction (\textbf{PVCP}) model for predicting the value of patents. This prediction aims to provide guidance for patent financing institutions.\par
Our main contributions are summarized as follows:\par
1.We have constructed a patent evaluation index system.\par
2.we have provided patent financing institutions with scientifically reliable patent valuation references.\par

\section{Related work}
\label{sect:Related_work}
\par
\subsection{Patent Application}
%早期专利分析方法不完善,只出现了一些定性分析方法和个别定量分析方法.在专利规划方面,Jeone和Yoon首次实现使用了专利定量分析方法提出了基于专利引用网络的新技术路线图,为专利技术提供了规划策略. 之后随着数据挖掘方法的完善,Chaoan和Cuilu结合数据挖掘和统计分析创建了基于可视化的专利挖掘分析框架用于发现智能家居的技术热点和专利空缺.专利信息化在专利检索方面的应用也取得了进展,Comins和Leydesdorff提出专利引文光谱法（PCS）,在生物医学研究与开发中构建了专利格局的一个基本组成部分是确定最具开创性的专利。其他专利信息化的应用还包括专利评估,这在接下来会详细介绍.
Early patent analysis methods were incomplete, with only a few qualitative and quantitative analysis methods emerging. In the field of patent planning, Jeone and Yoon pioneered the use of quantitative patent analysis methods to propose a new technological roadmap\citep{jeong2011technology} based on the patent citation network, providing strategic planning for patent technology. Subsequently, with the improvement of data mining methods\citep{agrawal1993mining}, Chaoan and Cuilu combined data mining and statistical analysis to create a visualization-based patent mining analysis framework\citep{chaoan2016application} for discovering technological hotspots and patent gaps in smart homes. Progress has also been made in the application of patent informatization in patent retrieval. Comins and Leydesdorff introduced the Patent Citation Spectrum (PCS) method, a fundamental component in constructing the patent landscape for biomedical research and development, aimed at identifying the most groundbreaking patents. Other applications of patent informatization include patent evaluation, which will be detailed in the following sections.\par
\subsection{Patent Evaluation}
%“专利价值”的概念首次由Sanders提出，他指出仅 55%专利被实施且各专利间的商业价值差异大。专利价值具有内外两个维度，内在价值主要指专利的技术价值如技术先进性、技术难度或复杂度等及授权后所具有的法律价值如权利要求数、同族数、被引证数等。外在价值主要指专利潜在市场能力，与专利市场价值、经济价值、新颖性、实用性、专利寿命等相关。下面从评估指标、评估方法两方面对专利价值评估进行梳理
The concept of "patent value" was first introduced by Sanders\citep{sanders}, who noted that only 55\% of patents are implemented, and there is a significant difference in the commercial value among patents. Patent value has two dimensions, internal and external. Internal value primarily refers to the technical aspects of a patent, such as technological advancement, technical complexity, and legal aspects like the number of claims, family size, and citation count after grant. External value pertains to a patent's potential marketability and its relevance to patent market value, economic value, novelty, practicality, patent lifespan, and other related factors. Subsequently, We will systematically review the evaluation of patent value from two aspects: evaluation indicators and evaluation methods.
%(1)评估指标.在最初的研究中,研究员采用单一指标专利价值评估方法,其中被引证数常使用[8]，该方法也受到质疑.兰德公司研究报告指出专利的被引用与其价值之间并非线性，二者之间的正相关性不理想。专利寿命[10]、专利授权率[11]、同族数[12]、权利要求数[13]等也常使用。该方法简单,但无法全面反映专利价值.之后研究员采用多指标综合评估,CHI Research 与美国国家科学基金会首次提出专利数量、被引证数、技术生命周期等 7 项经典的专利价值评估指标综合评估地区、企业整体的专利价值；Harhoff 等[14]选取专利范围、同族数等指标评估专利价值；Park[15]选取多个涉及技术内在特征和技术使用情况的相关指标构建专利价值评估指标体系.多因素指标组合是当前专利价值评估的主流方法，它降低单一指标评估的片面性和主观性，但依然面临指标设置冗余、指标权重设置不合理等问题。
Evaluation Indicators. In the initial studies, researchers adopted a single-indicator method for assessing patent value, with citation count being a common choice\citep{trajtenberg1990penny}. However, this method has faced scrutiny. A research report from the Rand Corporation indicated that the citation of patents does not exhibit a linear relationship with their value, and the positive correlation between the two is not ideal. Other commonly used indicators include patent lifespan\citep{schankerman1986estimates}, patent grant rate\citep{griliches1998patent}, family size\citep{neuhausler2013patent}, and number of claims\citep{llanes2012patent}. While this method is straightforward, it fails to comprehensively reflect the value of patents. Subsequently, researchers turned to a multi-indicator comprehensive assessment. CHI Research and the National Science Foundation in the United States proposed, for the first time, a comprehensive assessment of patent value for regions and overall enterprises, considering seven classical patent value evaluation indicators, including the quantity of patents, citation count, and technological lifecycle. Harhoff et al.\citep{harhoff2003citations} selected indicators such as patent scope and family size to evaluate patent value. Park\citep{park2004new} constructed a patent value evaluation indicator system by selecting multiple relevant indicators involving intrinsic technological features and technology usage. The combination of multiple factors is currently the mainstream method for patent value assessment, reducing the one-sidedness and subjectivity of single-indicator assessments, although it still faces challenges such as redundant indicator settings and unreasonable weightings.\par
%(2)评估方法.早期学者采用成本法,成本法并非由特定个人提出，而是由多个经济学家和评估专家在不同领域的研究中逐渐形成的。但是由于其忽略时间对专利价值的影响以及无法考虑市场需求和经济原理,该方法逐渐被淘汰.之后学者在成本法上改进并提出了数个方法,其中出名的是欧洲 IPScore 系统,但是这些方法都是定性描述专利价值,无法给出可靠的数据支持.20世纪80年代,随着Partial Least Squares(PLS)被引入到社会科学和管理研究领域, 专利价值指标得到了定量分析. Martinez Ruiz Alba和Aluja Banet Tomas在2009年提出了偏最小二乘 (PLS) 路径建模，将决定专利价值的变量联系起来，以可再生能源领域为例，其对变量相关性分析表现优异.之后,机器学习模型的兴起为专利价值研究带来了新的方法.Secil Ercan和Gulgun Kayakutlu使用SVM模型在家电家电专利行业构建出了智能分类模型，预测资助的可能性，帮助决策者预测专利上诉是否会被接受。
Evaluation Methods. In the early stages, scholars adopted the cost method, which was not proposed by a specific individual but gradually formed through the research of multiple economists and assessment experts in various fields. However, due to its neglect of the impact of time on patent value and the inability to consider market demand and economic principles, this method was gradually phased out. Subsequently, scholars improved and proposed several methods based on the cost method, with the notable European IPScore system among them. However, these methods only qualitatively describe patent value and cannot provide reliable data support.In the 1980s, with the introduction of Partial Least Squares (PLS)\citep{wold1975soft} into the fields of social science and management research, patent value indicators underwent quantitative analysis. In 2009, Martinez Ruiz Alba and Aluja Banet Tomas proposed the PLS path modeling, connecting variables that determine patent value. They demonstrated excellent performance in correlational analysis, taking the renewable energy sector as an example.Subsequently, the rise of machine learning models brought new approaches to patent value research. Secil Ercan and Gulgun Kayakutlu used the Support Vector Machine(SVM) model\citep{cortes1995support} to construct an intelligent classification model in the household appliance patent industry\citep{ercan2014patent}, predicting the likelihood of funding and assisting decision-makers in anticipating whether a patent appeal would be accepted.\par

\section{Basic Theory and Research Design}
\label{sect:Basic_Theory_and_Research_Design}
\par
\subsection{Basic Theory and Methods}
\subsubsection{Traditional Methods}
%描述性统计分析。这种方法的主要目的是通过汇总和分析数据集的基本统计信息来更好地理解数据的分布和特征。统计信息包括均值、中位数、方差、标准差、分位数等信息。
Descriptive statistical analysis. The main purpose of this method is to better understand the distribution and characteristics of the data by summarizing and analyzing basic statistical information about the data set. Statistical information includes information such as mean, median, variance, standard deviation, and quantile.\par
%ANOVA。
ANOVA. ANOVA is a statistical test used to compare differences in means between three or more groups (or samples). Its main purpose is to determine whether at least one group has a mean that is different from the others.\par
%皮尔逊相关系数。皮尔逊相关系数是一种用于衡量两个变量之间线性关系的统计方法。该方法通过计算相关系数来评估这两个变量之间的相关程度，相关系数的值介于-1和1之间，反映了相关性的方向和强度。皮尔逊相关系数通过除以两个变量的标准差的乘积来标准化协方差，从而得到一个不受单位影响的度量。这使得 r 能够量化两个变量的线性关系的强度和方向。
Pearson correlation coefficient. Pearson's correlation coefficient is a statistical method used to measure the linear relationship between two variables. The method assesses the degree of correlation between these two variables by calculating the correlation coefficient, which has a value between -1 and 1, reflecting the direction and strength of the correlation. The Pearson correlation coefficient standardizes the covariance by dividing it by the product of the standard deviations of the two variables, resulting in a unit-independent measure. This allows r to quantify the strength and direction of the linear relationship between two variables.\par
\subsubsection{AI Methods}
%决策树分类算法。决策树分类算法是一种常见的监督学习算法，用于解决分类问题。它的原理基于树状结构，通过递归地选择最佳特征进行数据分割，从而构建一棵决策树，该树用于对新样本进行分类。决策树分类算法的优点之一是其可解释性强。生成的决策树可以直观地展示每个决策节点所基于的特征和阈值，使决策过程易于理解。此外，决策树适用于各种数据类型，包括分类特征和数值特征。另一个优点是它在处理大型数据集时具有较高的计算效率。
DTC. DTC algorithm is a common supervised learning algorithm for solving classification problems. It is based on the principle of tree structure, where data is partitioned by recursively selecting the best features to construct a decision tree which is used to classify new samples. One of the advantages of the DTC algorithm is its interpretability. The generated decision tree visualizes the features and thresholds on which each decision node is based, making the decision-making process easy to understand. In addition, decision trees are suitable for a variety of data types, including categorical and numerical features. Another advantage is its high computational efficiency when dealing with large datasets.\par
\subsection{Research Design}
\subsubsection{Patent Data Collection}
%本研究的专利数据由电梯行业制造商提供.我们一共收集到252047条数据、其中每条专利信息包含50个指标信息。
This study's patent data was provided by manufacturers in the elevator industry. In total, we collected 252,047 records, with each patent information containing 50 indicators.\par
\subsubsection{Feature Selection}
%参考以往学者的成果以及我国《“十四五”国家知识产权保护和运用规划》政策，我们依据科学性和易获取性选取了15个专利指标。分别为：产业链位置、一级技术分支、专利类型、公开国别、公开(公告)日、权利要求数量、文献页数、IPC、申请人省市代码、3年内被引用次数、5年内被引用次数、引用专利数量、被引用专利数量、专利有效性、诉讼案件数。
Referring to the achievements of previous scholars and the policies outlined in China's "14th Five-Year National Plan for Intellectual Property Protection and Utilization," we selected 15 patent indicators based on their scientific relevance and availability. These indicators include: Industry Chain Position, Primary Technical Branch, Patent Type, Publication Country, Publication Date, Number of Claims, Number of Document Pages, IPC (International Patent Classification), Applicant's Province and City Code, Number of Citations within 3 Years, Number of Citations within 5 Years, Number of Citing Patents, Number of Cited Patents, Patent Validity, and Number of Litigation Cases, Patent Lifespan. \par
%其中分类指标为：产业链位置、一级技术分支、专利类型、公开国别、IPC、申请人省市代码、专利有效性。定量指标为：公开(公告)日、权利要求数量、文献页数、3年内被引用次数、5年内被引用次数、引用专利数量、被引用专利数量、诉讼案件数。
Among these indicators, the categorical ones are Industry Chain Position, Primary Technical Branch, Patent Type, Publication Country, IPC, Applicant's Province and City Code, and Patent Validity. The quantitative indicators include Publication Date, Number of Claims, Number of Document Pages, Number of Citations within 3 Years, Number of Citations within 5 Years, Number of Citing Patents, Number of Cited Patents, and Number of Litigation Cases, Patent Lifespan.\par
\subsubsection{Feature and Correlation Analysis}
%对于分类指标，我们首先统计专利价值在各个指标上的均值、中位数、方差等信息，使用ANOVA分析专利价值在各个指标上的差异性。对于定量指标，我们使用皮尔逊相关系数分析两两指标之间的相关程度。通过以上两个方面得到每个指标和专利价值的相关程度。
For categorical indicators, we first calculate the mean, median, variance, and other statistics of patent values within each indicator. We use ANOVA to analyze the differences in patent values among these indicators. For quantitative indicators, we employ Pearson correlation coefficients to analyze the relationships between pairs of indicators. Through these two approaches, we obtain the degree of correlation between each indicator and patent value.\par
\subsubsection{Construction of PVCP Model}
%我们使用以上15个指标作为模型特征。因为决策树分类模型具有很好的解释性，以及对大数据集有很好的运算效率，我们选择决策树分类模型作为训练模型。为保证结果的准确率，我们使用了交叉检验的方式进行模型训练。
We used the aforementioned 15 indicators as model features. Due to the DTC model's strong interpretability and efficient processing of large datasets, we chose it as our training model. To ensure the accuracy of the results, we employed cross-validation\cite{box1986analysis} during the model training process.\par
\section{Experimental Implementation}
\label{sect:Experimental_Implementation}
\par
\subsection{Data Preprocessing}
%我们将以上15个指标中值为空的数据点删除，最终得到72037条数据，其中专利寿命指标为空的数据点超过50%，所以我们将专利寿命指标删去，最终保留包含14个指标的72037条数据。
After removing data points with missing values for the selected 15 indicators, we obtained a final dataset of 72,037 records. As more than 50\% of the data points had missing values for the patent lifespan indicator, we decided to exclude this indicator. In the end, we retained a dataset comprising 72,037 records with 14 indicators.\par
\begin{figure}[H]
  \centering
  \includegraphics[width=1.0\linewidth]{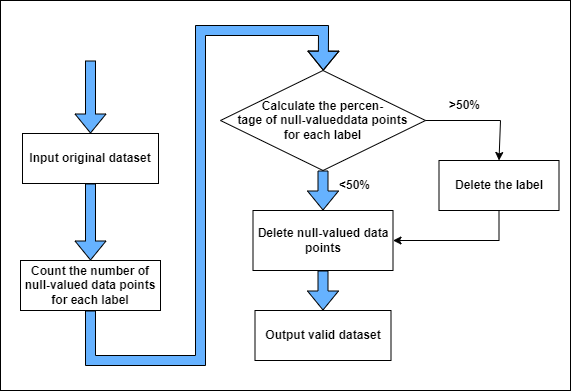} 
  \caption{Data preprocessing process}
  \label{fig:dataprocess}
\end{figure}
\subsection{Global Patent Analysis}
%1. Publication Country。我们统计了每个国家的电梯专利数量和均值。因为国家数量众多，这里我们以中国、美国、日本、韩国、德国为主要目标进行统计，结果如图1。
1. Publication Country. We have compiled the number of elevator patents for each country. Given the numerous countries, we primarily focused on China, the United States, Japan, South Korea, and Germany for our analysis. The results are shown in Figure 1.\par
% \begin{figure}[h]
%     \centering
%     \includegraphics[width=0.5\textwidth]{country_number_Analysis.png}
%     \caption{Distribution of elevator patents by country}
% \end{figure}

\begin{figure}[H]
    \begin{subfigure}{0.5\textwidth}
        \includegraphics[width=\linewidth]{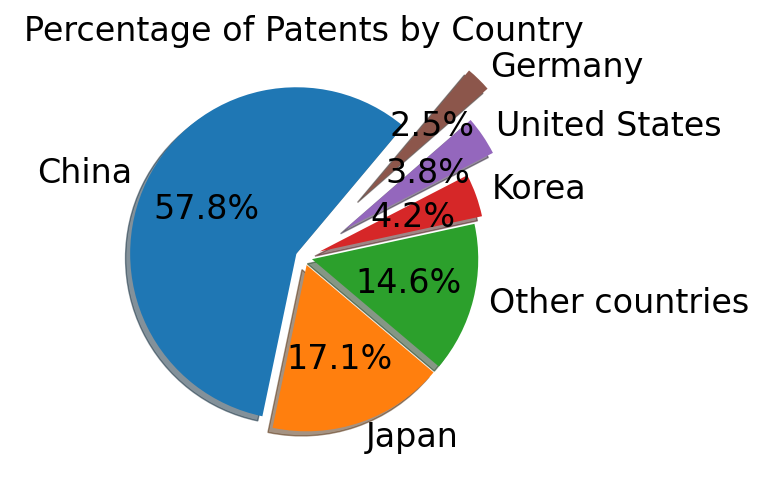}
        \caption{Proportion of Patents by Country}
        \label{fig:country_number_Analysis}
    \end{subfigure}
    \begin{subfigure}{0.5\textwidth}
        \includegraphics[width=\linewidth]{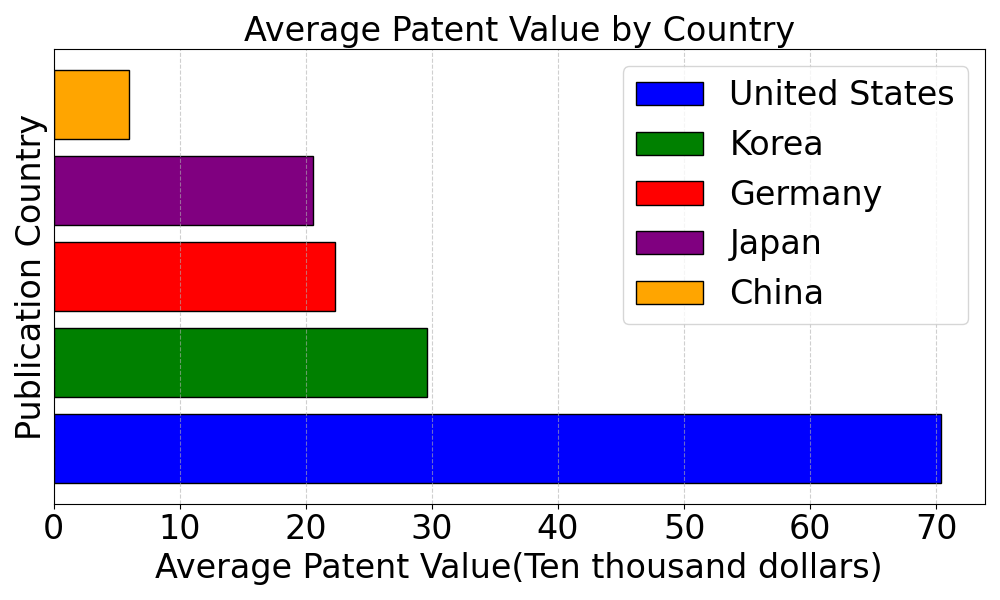}
        \caption{Average patent value by country}
        \label{fig:Country_mean_analysis}
    \end{subfigure}
    \caption{Distribution of the Proportion and Average Value of Patents by Country}
\end{figure}
%从图中得到我国在电梯领域中专利数量所占的比例为57.8%，日本为37.1%，韩国为4.2%，美国为3.8%，德国为2.5%，其他国家一共占到14.6%，可见专利数量在不同国家有差异性。除此之外，专利价值均值从小到大一次是：中国、日本、德国、韩国、美国。图中所展示的国外专利价值比国内专利价值普遍要高的现象恰好由国外专利价值申请成本高于国内的原因所解释。同时可见专利价值在不同国家有差异性。综上：Publication Country是相关特征。
From the chart, we can observe that China accounts for 57.8\% of the total number of patents in the elevator industry, followed by Japan at 37.1\%, South Korea at 4.2\%, the United States at 3.8\%, and Germany at 2.5\%. Other countries collectively make up 14.6\%, indicating variations in patent numbers among different nations. Furthermore, the mean patent value ranks as follows, from lowest to highest: China, Japan, Germany, South Korea, and the United States. This reflects the common trend of foreign patents having higher values, primarily due to the higher costs associated with applying for foreign patents compared to domestic ones. It is evident that patent values also vary across different countries. In conclusion, 'Publication Country' is a relevant feature.\par

\subsection{Chinese Patent Analysis}

\subsubsection{Analysis of Categorical Features}

%2.Applicant's Province and City Code.我们统计了中国每个省份的专利数量占比和专利价值均值，如图2.
2. Applicant's Province and City Code. We calculated the proportion of patent numbers and the mean patent value for each province in China, as shown in Figure 2.\par
\begin{figure}[H]
    \begin{subfigure}{0.5\textwidth}
        \includegraphics[width=\linewidth]{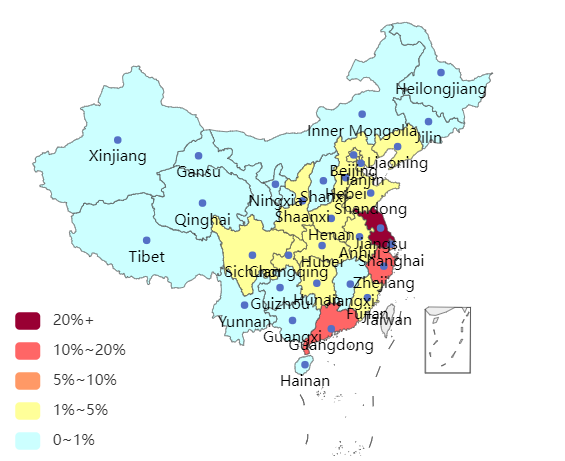}
        \caption{Proportion of Patents by Province of China}
    \end{subfigure}
    \begin{subfigure}{0.5\textwidth}
        \includegraphics[width=\linewidth]{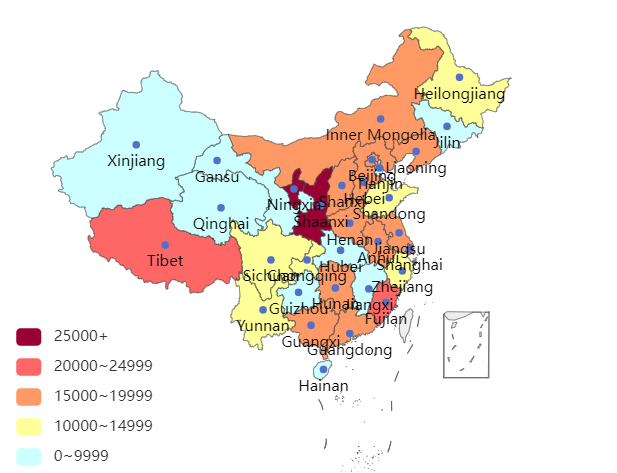}
        \caption{Average Patent Value by Province of China}
    \end{subfigure}
    \caption{Proportion and Average Value of Patents by Province of China}
\end{figure}
%从图中发现沿海省份的专利数量占全国75%以上，专利价值均值也是沿海省份和靠近海洋的内陆省份居高，说明沿海地区在科技专利上有一定的优势，也说明Applicant's Province and City Code与专利价值有相关性。
From the chart, it is evident that coastal provinces account for over 75\% of the total patent quantity in the entire country. Additionally, the mean patent value is higher in these coastal provinces as well as in inland provinces that are closer to the sea. This indicates that coastal regions have a certain technological advantage in patents, and it further suggests a correlation between 'Applicant's Province and City Code' and patent value.\par
%4.Industry Chain Position, Primary Technical Branch, Patent Type,IPC,Patent Validity.这些指标为一般的分类特征，图2-6为专利价值分别在这些指标下的均值分布。
%3.Publication Date.Publication Date.是时间属性，既可以作为分类特征也可以作为定量特征，我们将两种类型都进行分析。这里我们首先分析分类特征，如图3.
3. Publication Date.Publication Date. Publication Date is a temporal attribute that can be analyzed as both a categorical and a quantitative feature. In this section, we will begin by examining its role as a categorical feature, as depicted in Figure 3.\par
%从图中可以发现自2003年-2009年，每年的专利价值均值有上下波动，但是从2009年以来，专利价值均值逐年下降，并且从总体上看专利价值均值也在下降。专利价值在不同年份上波动较大，反映了年份与专利价值存在相关性。
The chart reveals that from 2003 to 2009, the annual average of patent values fluctuated up and down. However, since 2009, the annual average of patent values has been decreasing year by year. Overall, the average patent value is also declining. The significant fluctuations in patent values in different years reflect a correlation between the year and patent value.\par
\begin{figure}[H]
    \centering
    \includegraphics[width=\textwidth]{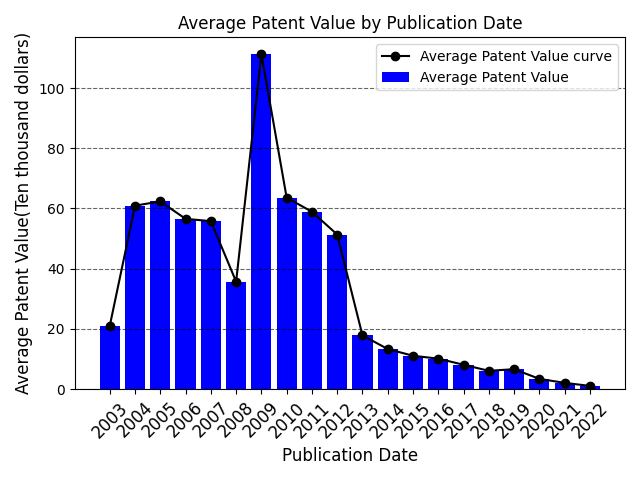}
    \caption{Average Patent Value by Publication Date}
    \label{fig:date_mean_analysis}
\end{figure}
4. Industry Chain Position, Primary Technical Branch, Patent Type,IPC,Patent Validity. These indicators are typical categorical features, and Figures 4-6 display the mean distribution of patent values within these indicators.\par
%从图2-6可以发现，专利价值在Primary Technical Branch上的极差在3万美元左右，在IPC上的极差在2万美元左右，在Patent Type上的极差在25万美元左右，在Industry Chain Position上的极差在3.5万美元左右，在Patent Validity上的极差在4.5万美元左右。使用ANOVA分析，发现不存在均值相同的分组，可见专利价值在以上不同指标下存在差异，即这些指标与专利价值相关。
From Figures 4 to 6, it can be observed that the range of patent values within the "Primary Technical Branch" indicator is approximately \$30,000, within "IPC" it is around \$20,000, within "Patent Type" it's roughly \$250,000, within "Industry Chain Position" it's about \$35,000, and within "Patent Validity" it's approximately \$45,000. Through ANOVA analysis, it was determined that there are no groups with the same means, indicating differences in patent values across the various indicators. Therefore, it can be concluded that these indicators are related to patent value.\par
\begin{figure}[H]
    \centering
    \includegraphics[width=0.8\textwidth]{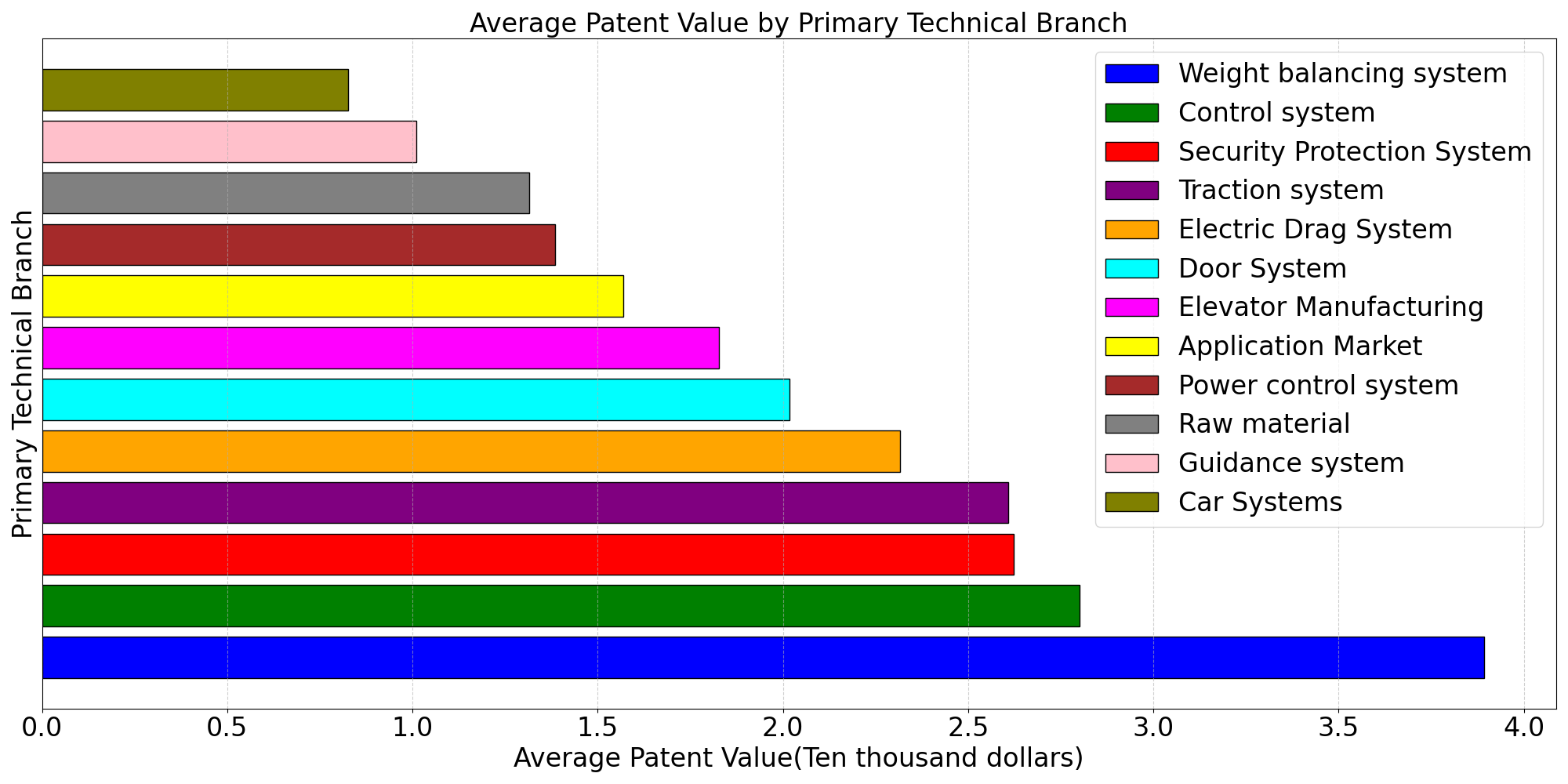}
    \caption{Average Patent Value by Primary Technical Branch}
    \label{fig:PrimaryTechnicalBranch_mean_analysis}
\end{figure}
\begin{figure}[H]
    \centering
    \includegraphics[width=0.8\textwidth]{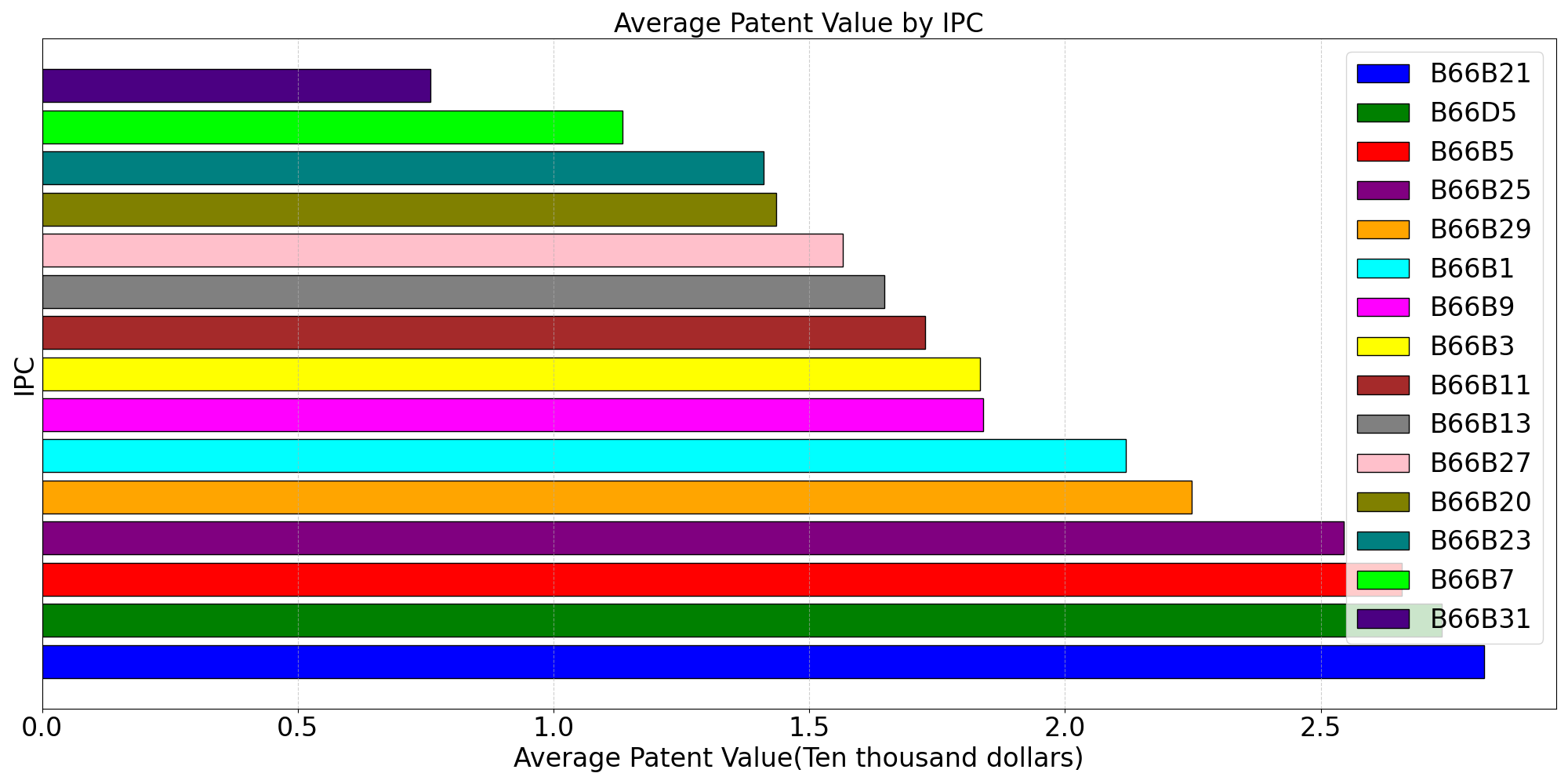}
    \caption{Average Patent Value by IPC}
    \label{fig:IPC_mean_analysis}
\end{figure}
\begin{figure}[H]
    \begin{minipage}{0.5\textwidth}
        \centering
        \includegraphics[width=\textwidth]{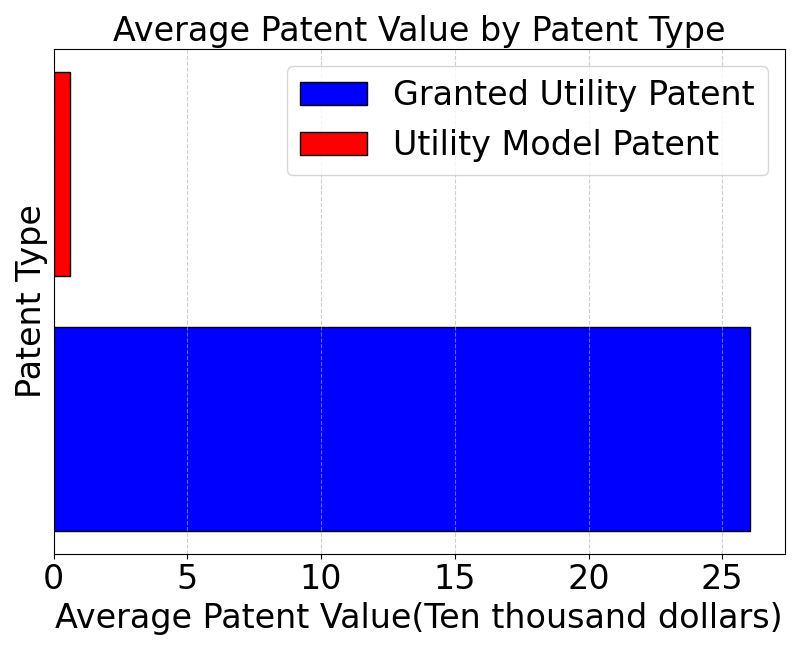}
        \caption{Average Patent Value by Patent Type}
        \label{type_mean_analysis}
    \end{minipage}\hfill
    \begin{minipage}{0.5\textwidth}
        \centering
        \includegraphics[width=\textwidth]{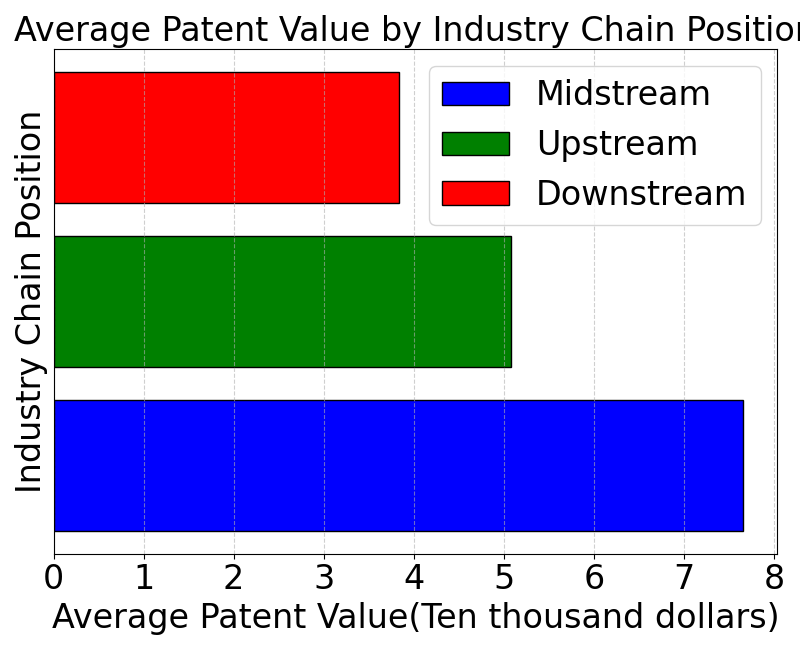}
        \caption{Average Patent Value by Industry Chain Position}
        \label{Chain_analysis}
    \end{minipage}\hfill
\end{figure}
\begin{figure}[H]
    \centering
    \includegraphics[width=0.8\textwidth]{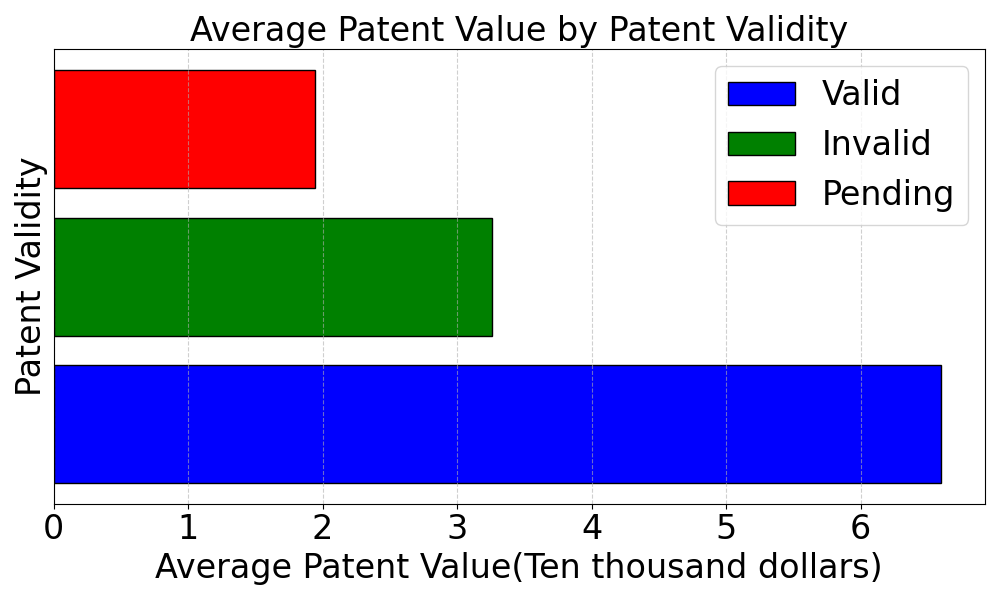}
    \caption{Average Patent Value by Validity}
    \label{fig:Validity_mean_analysis}
\end{figure}
\subsubsection{Analysis of Quantitative Features}
%皮尔逊相关系数计算公式：
Pearson correlation coefficient calculation formula:
$$\rho = \frac{\sum \left ( x_i-\bar{x}   \right )\left ( y_i-\bar{y}  \right )  }{\sqrt{\sum \left ( x_i-\bar{x}  \right )^{2}\sum \left ( y_i-\bar{y}  \right ) ^{2}   } } $$
%其中x_i和y_i是是样本中的个体数据点;\bar{x}和\bar{y}是 x 和 y 的均值(平均值);符号\sum表示对所有数据点进行求和操作.
Where $x_i$ and $y_i$ are individual data points in the sample; $\bar{x}$ and $\bar{y}$ are the means of x and y;the symbol $\sum$ represents the summation over all data points.\par
%我们对定量特征两两进行皮尔逊相关性分析，使用编程进行相关系数的计算，这里以Patent Value和Number of Claims为例，以Patent Value作为x，Number of Claims作为y。首先我们将数据集中的这两个属性提取出来，然后求出Patent Value和Number of Claims的均值，即\bar{x},\bar{y},然后将数据点输入公式中求取皮尔逊相关系数。皮尔逊相关系数矩阵见图8.
We perform pairwise Pearson correlation analysis on quantitative features and calculate correlation coefficients programmatically. Here, we take Patent Value and Number of Claims as an example, with Patent Value as x and Number of Claims as y. First, we extract these two attributes from the dataset. Then, we calculate the means of Patent Value and Number of Claims, denoted as $\bar{x}$ and $\bar{y}$. Subsequently, we input the data points into the formula to calculate the Pearson correlation coefficient.The Pearson correlation coefficient matrix is shown in Figure 9.\par
\begin{figure}[H]
    \centering
    \includegraphics[width=\textwidth]{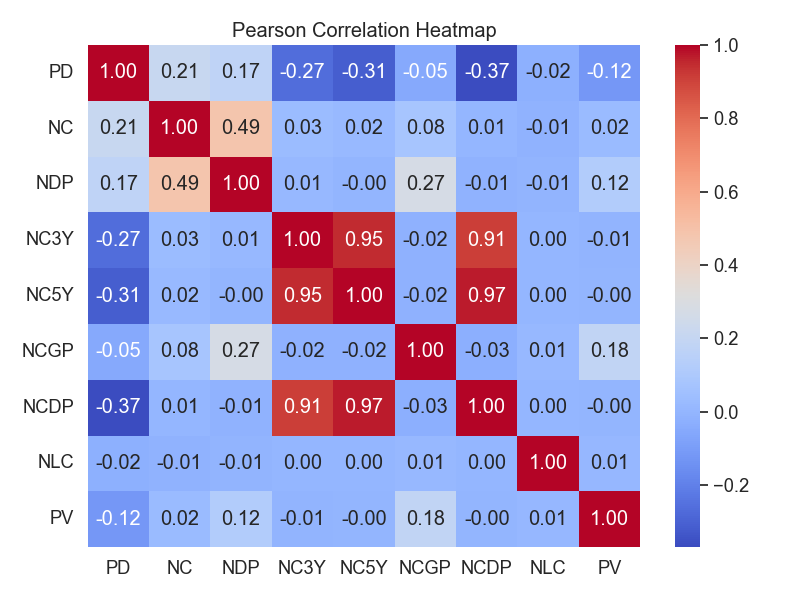}
    \caption{Pearson Correlation Coefficient Heatmap}
\end{figure}
%由于标签名字过长，我们采用标签各单词的首字母作为缩简标签。图中PD(Publication Date)与PV(Patent Value)成负相关关系，这与我们之前将Publication Date作为分类特征得到的统计结果一致。NDP(Number of Document Pages)与Patent Value成一定正相关性。重要的是, 我们发现NCGP(Number of Citing Patents)与Patent Value有一定相关性而NCDP(Number of Cited Patents)与Patent Value几乎没有相关性。在其他关于专利价值的研究中，普遍认为一个专利被引用的次数越多，该专利越有影响力，其价值应当越高，但是根据数据显示，一个专利引用别人专利越多，价值越高。
Due to the lengthy label names, we have abbreviated them using the initials of each word. In the figure, PD (Publication Date) shows a negative correlation with PV (Patent Value), which aligns with the statistical results we obtained when treating Publication Date as a categorical feature. NDP (Number of Document Pages) exhibits a moderate positive correlation with Patent Value. Importantly, we observed that NCGP (Number of Citing Patents) is moderately correlated with Patent Value, while NCDP (Number of Cited Patents) shows almost no correlation with Patent Value. In most studies on patent value, it is commonly believed that the more a patent is cited, the more influential and valuable it should be. However, based on our data, a patent's value tends to increase as it cites other patents more frequently.\par
\subsection{Construction of PVCP Model}
%我们首先将数据集中的专利数据按照专利价值进行分类，添加“专利价值分类”标签，以一万美元以上为高价值，一万美元以下为低价值进行填充。然后对数据进行空值清洗。最后使用决策树分类模型进行模型训练，分别求出不同深度下的叶子节点和准确率信息。
We first classified the patent data in the dataset based on patent value, adding a 'Patent Value Classification' label, with values filled as 'large' for patents valued above \$10,000 and 'small' for patents valued at or below \$10,000. We then performed data cleaning to handle missing values. Finally, we trained a decision tree classification model and calculated information about the number of leaf nodes and accuracy for different depths.\par
%表1是决策树深度与测试集的正确率的对应关系
Table 1 illustrates the relationship between decision tree depth and the accuracy on the test set.\par
\begin{table}[H]
\centering
\caption{the Accuracy of DTC Models with Different Tree Depths}
\begin{tabular}{|c|c|}
\hline
Accuracy Max layer & Accuracy \\ \hline
\hline
1                  & 96.65\%  \\ \hline
2                  & 97.92\%  \\ \hline
3                  & 98.39\%  \\ \hline
4                  & 98.39\%  \\ \hline
5                  & 98.48\%  \\ \hline
\end{tabular}
\end{table}
%结果分析：从表1中可以看出该决策树的表现优异，在树的深度达到第11层时，模型的准确率达到99%，但是出现了一个不正常的现象，决策树在第一层决策时达到了96.65%的正确率。为分析这一现象，我们做出了前三层决策树过程图，如图10。
Results Analysis: From Table 1, it can be observed that the decision tree performs exceptionally well, achieving an accuracy of 99\% when the tree reaches a depth of 10 layers. However, there is an unusual phenomenon where the decision tree attains a 96.65\% accuracy at the first-level decision. To analyze this phenomenon, we have created process diagrams for the first three levels of the decision tree, as shown in Figure 10.\par
\begin{figure}[H]
    \centering
    \includegraphics[width=\textwidth]{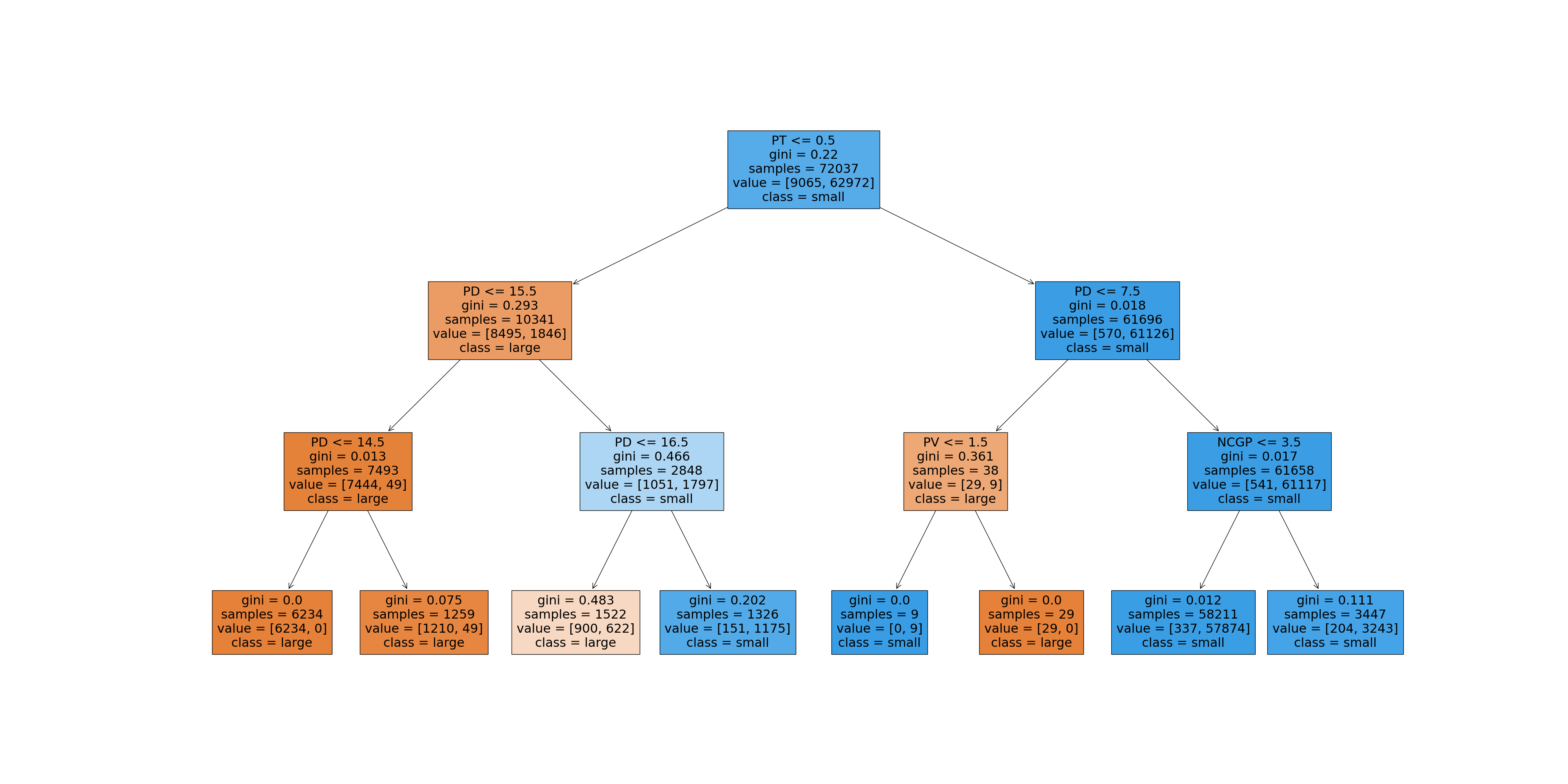}
    \caption{Decision Tree Process Diagram}
\end{figure}
%从图10中发现，决策树第一层将PT(Patent Type)作为判断条件，于是我们统计了small value和large value在不同Patent Type下的占比，如表2
From Figure 10, it was observed that the decision tree uses 'Patent Type (PT)' as a decision criterion at the first level. Therefore, we calculated the percentages of 'small value' and 'large value' under different 'Patent Type' categories, as shown in Table 2.\par
\begin{table}[H]
\centering
\caption{the Accuracy of DTC Models with Different Tree Depths}
\begin{tabular}{|c|c|c|}
\hline
            & Granted Utility Patent & Utility Model Patent \\ \hline
small value & 3.65\%                 & 96.35\%              \\ \hline
large value & 98.44\%                & 1.56\%               \\ \hline
\end{tabular}
\end{table}

\label{}

\section{Conclusion}
\label{sect:Conclusion}








% \end{thebibliography}
\bibliography{BibTex}

\begin{thebibliography}{16}
\providecommand{\natexlab}[1]{#1}
\providecommand{\url}[1]{\texttt{#1}}
\expandafter\ifx\csname urlstyle\endcsname\relax
  \providecommand{\doi}[1]{doi: #1}\else
  \providecommand{\doi}{doi: \begingroup \urlstyle{rm}\Url}\fi

\bibitem[Quinlan(1986)]{quinlan1986induction}
J.~Ross Quinlan.
\newblock Induction of decision trees.
\newblock \emph{Machine learning}, 1:\penalty0 81--106, 1986.
\newblock \doi{https://doi.org/10.1007/BF00116251}.

\bibitem[Jeong and Yoon(2011)]{jeong2011technology}
Yu-Jin Jeong and Byung-Un Yoon.
\newblock Technology planning through technology roadmap: Application of patent citation network.
\newblock \emph{Journal of the Korea Academia-Industrial Cooperation Society}, 12\penalty0 (11):\penalty0 5227--5237, 2011.
\newblock \doi{https://doi.org/10.5762/KAIS.2011.12.11.5227}.

\bibitem[Agrawal et~al.(1993)Agrawal, Imieli{\'n}ski, and Swami]{agrawal1993mining}
Rakesh Agrawal, Tomasz Imieli{\'n}ski, and Arun Swami.
\newblock Mining association rules between sets of items in large databases.
\newblock In \emph{Proceedings of the 1993 ACM SIGMOD international conference on Management of data}, pages 207--216, 1993.
\newblock \doi{https://doi.org/10.1145/170035.170072}.

\bibitem[Chaoan and Cuilu(2016)]{chaoan2016application}
LAI Chaoan and XU~Cuilu.
\newblock The application of patent mining in the forecast of smart home industry.
\newblock \emph{Management Science and Engineering}, 10\penalty0 (1):\penalty0 67--75, 2016.
\newblock \doi{https://doi.org/10.3968/8220}.

\bibitem[SANDERS B~S(1958)]{sanders}
HARRIS L~J SANDERS B~S, ROSSMAN~J.
\newblock The economic impact of patents[j].
\newblock \emph{Trademark and Copyright Journal}, 2\penalty0 (2), 1958.

\bibitem[Trajtenberg(1990)]{trajtenberg1990penny}
Manuel Trajtenberg.
\newblock A penny for your quotes: patent citations and the value of innovations.
\newblock \emph{The Rand journal of economics}, pages 172--187, 1990.
\newblock \doi{https://doi.org/10.2307/2555502}.

\bibitem[Schankerman and Pakes(1986)]{schankerman1986estimates}
Mark Schankerman and Ariel Pakes.
\newblock Estimates of the value of patent rights in european countries during the post-1950 period.
\newblock \emph{The economic journal}, 96\penalty0 (384):\penalty0 1052--1076, 1986.
\newblock \doi{https://doi.org/10.2307/2233173}.

\bibitem[Griliches(1998)]{griliches1998patent}
Zvi Griliches.
\newblock Patent statistics as economic indicators: a survey.
\newblock In \emph{R\&D and productivity: the econometric evidence}, pages 287--343. University of Chicago Press, 1998.

\bibitem[Neuh{\"a}usler and Frietsch(2013)]{neuhausler2013patent}
Peter Neuh{\"a}usler and Rainer Frietsch.
\newblock Patent families as macro level patent value indicators: applying weights to account for market differences.
\newblock \emph{Scientometrics}, 96:\penalty0 27--49, 2013.
\newblock \doi{https://doi.org/10.1007/s11192-012-0870-y}.

\bibitem[Llanes and Trento(2012)]{llanes2012patent}
Gaston Llanes and Stefano Trento.
\newblock Patent policy, patent pools, and the accumulation of claims in sequential innovation.
\newblock \emph{Economic Theory}, 50:\penalty0 703--725, 2012.
\newblock \doi{https://doi.org/10.1007/s00199-010-0591-5}.

\bibitem[Harhoff et~al.(2003)Harhoff, Scherer, and Vopel]{harhoff2003citations}
Dietmar Harhoff, Frederic~M Scherer, and Katrin Vopel.
\newblock Citations, family size, opposition and the value of patent rights.
\newblock \emph{Research policy}, 32\penalty0 (8):\penalty0 1343--1363, 2003.
\newblock \doi{https://doi.org/10.1016/S0048-7333(02)00124-5}.

\bibitem[Park and Park(2004)]{park2004new}
Yongtae Park and Gwangman Park.
\newblock A new method for technology valuation in monetary value: procedure and application.
\newblock \emph{Technovation}, 24\penalty0 (5):\penalty0 387--394, 2004.
\newblock \doi{https://doi.org/10.1016/S0166-4972(02)00099-8}.

\bibitem[Wold(1975)]{wold1975soft}
Herman Wold.
\newblock Soft modelling by latent variables: the non-linear iterative partial least squares (nipals) approach.
\newblock \emph{Journal of Applied Probability}, 12\penalty0 (S1):\penalty0 117--142, 1975.
\newblock \doi{https://doi.org/10.1017/S0021900200047604}.

\bibitem[Cortes and Vapnik(1995)]{cortes1995support}
Corinna Cortes and Vladimir Vapnik.
\newblock Support-vector networks.
\newblock \emph{Machine learning}, 20:\penalty0 273--297, 1995.
\newblock \doi{https://doi.org/10.1007/BF00994018}.

\bibitem[Ercan and Kayakutlu(2014)]{ercan2014patent}
Secil Ercan and Gulgun Kayakutlu.
\newblock Patent value analysis using support vector machines.
\newblock \emph{Soft computing}, 18\penalty0 (2):\penalty0 313--328, 2014.
\newblock \doi{https://doi.org/10.1007/s00500-013-1059-x}.

\bibitem[Box and Meyer(1986)]{box1986analysis}
George~EP Box and R~Daniel Meyer.
\newblock An analysis for unreplicated fractional factorials.
\newblock \emph{Technometrics}, 28:\penalty0 11--18, 1986.
\newblock \doi{https://doi.org/10.1080/00401706.1986.10488093}.

\end{thebibliography}
% \bibliography{references}
\end{document}